\documentclass[aps,pre,twocolumn,superscriptaddress,10pt]{revtex4-1}
\usepackage{amsmath,newtxtext,newtxmath,bm}
\usepackage[pdftex]{graphicx} 

\newcommand{\sref}[1]{Sect.~\ref{#1}}
\newcommand{\eref}[1]{Eq.~\eqref{#1}}
\newcommand{\fref}[1]{Fig.~\ref{#1}}
\newcommand{\aref}[1]{Appendix~\ref{#1}}
\DeclareMathOperator*{\argmin}{arg\,min}

\DeclareMathOperator*{\extr}{extr}

\begin{document}
\title{Statistical-mechanical analysis of compressed sensing for Hamiltonian estimation of Ising spin glass}
\author{Chako Takahashi}
\email{chako@smapip.is.tohoku.ac.jp}
\affiliation{Graduate School of Information Sciences, Tohoku University, 6-3-09 Aramaki-aza-Aoba, Aoba-ku, Sendai, Miyagi 980-8579, Japan}
\author{Masayuki Ohzeki}
\email{mohzeki@tohoku.ac.jp}
\affiliation{Graduate School of Information Sciences, Tohoku University, 6-3-09 Aramaki-aza-Aoba, Aoba-ku, Sendai, Miyagi 980-8579, Japan}
\author{Shuntaro Okada}
\affiliation{Graduate School of Information Sciences, Tohoku University, 6-3-09 Aramaki-aza-Aoba, Aoba-ku, Sendai, Miyagi 980-8579, Japan}
\affiliation{Advanced Research and Innovation Division 3, DENSO CORPORATION, 1-1 Showa-cho, Kariya, Aichi 448-8661, Japan}
\author{Masayoshi Terabe}
\affiliation{Advanced Research and Innovation Division 3, DENSO CORPORATION, 1-1 Showa-cho, Kariya, Aichi 448-8661, Japan}
\author{Shinichiro Taguchi}
\affiliation{Advanced Research and Innovation Division 3, DENSO CORPORATION, 1-1 Showa-cho, Kariya, Aichi 448-8661, Japan}
\author{Kazuyuki Tanaka}
\affiliation{Graduate School of Information Sciences, Tohoku University, 6-3-09 Aramaki-aza-Aoba, Aoba-ku, Sendai, Miyagi 980-8579, Japan}

\begin{abstract}
Several powerful machines, such as the D-Wave 2000Q, dedicated to solving combinatorial optimization problems through the Ising-model formulation have been developed.
To input problems into the machines, the unknown parameters on the Ising model must be determined, and this is necessarily a nontrivial task.
It could be beneficial to construct a method to estimate the parameters of the Ising model from several pairs of values of the energy and spin configurations.
In the present paper, we propose a simple method employing the $L_1$-norm minimization, which is based on the concept of the compressed sensing.
Moreover, we analyze the typical performance of our proposed method of the Hamiltonian estimation by using the replica method.
We also compare our analytical results through several numerical experiments using the alternating direction method of multipliers.
\end{abstract}

\maketitle

\section{Introduction}
\label{sec:intro}

Recent years have shown a rapid increase in the development of new types of computers.
A representative of such new types of computers is the D-Wave machine (current system is called D-Wave 2000Q), which implements quantum annealing~\cite{Dwave2011}.
Quantum annealing is invented for solving the optimization problem by utilizing the quantum fluctuation~\cite{Kadowaki1998,Morita2008,Ohzeki2011c}.
In ideal procedure of quantum annealing, the quantum adiabatic theorem assures slow control of the quantum fluctuation fixes time-evolved state into nontrivial ground state of the instantaneous Hamiltonian.
The computational time to attain the ground state is characterized by the energy gap between the ground and excited states~\cite{Suzuki2005}.
Beyond the scheme of the adiabatic quantum theorem, non-adiabatic approaches are also proposed in several ways~\cite{Ohzeki2010a,Ohzeki2011,Ohzeki2011proc,Somma2012}.
Recent analysis reveals possibility of exponential speedup by utilizing the relaxation of the excited states into the ground state due to the thermal effect.
In addition, various attempts beyond the standard quantum fluctuations are demonstrated to attain quantum speedup by implementing some tricky quantum effects~\cite{Seki2012,Seki2015,Ohzeki2017,Nishimori2017,Okada2018}.
In order to investigate the state of the spin configuration after quantum annealing, a machine-learning technique is also proposed~\cite{Arai2018}.
The D-Wave machine can solve combinatorial optimization problems and speedily perform the Gibbs--Boltzmann sampling.
The D-Wave machine implements the Ising model on its superconducting circuit~\cite{Dwave2010b,Dwave2010c,Dwave2014}.
Thus, the optimization problems are written in terms of the Ising Hamiltonian and restricted to the case only with biases and two-body interactions on the peculiar sparse graph, known as the chimera graph, and bias terms.
When the type of Ising Hamiltonian of the optimization problem is already known, the problem can be directly implemented in the D-Wave machine.
Beyond the chimera graph in the D-Wave machine, the generic form of the Ising Hamiltonian is fully-connected model, namely, the interactions are dense.
We then employ a specialized technique, which has been frequently studied, to embed the Ising Hamiltonian on the D-Wave machine.
However, in most of the cases, we do not necessarily know how to formulate the optimization problem in terms of Ising Hamiltonian.
In our problem setting, we do not know the detailed form of the cost function but only the pairs of the spin configurations and energy values of the Ising Hamiltonian.
Here we assume that it takes a relatively long time to attain the value of the cost function on the given input.
For unfamiliar readers, we take one of the representative.
When we investigate the effect of some drugs, it appears after long time duration.
We cannot attain immediate response of the combination of the selected drug.
Then we infer the cost function only from several pairs of input and output.

This is a typical inverse problem to infer an Ising Hamiltonian from pairs of the input and output because the Hamiltonian can be regarded as a function that outputs the energy value from the spin configuration as the input.
We then assume that the Ising Hamiltonian has sparse interactions, and thus can be rather easily embedded on the D-Wave machine.
Next, the inverse problem considered in the present study is formulated as the setting of the so-called compressed sensing (CS) in terms of the signal processing~\cite{Donoho2006}.
In CS, we aim at reconstruction of the original signal from a small number of observations.
In general, the problem for recovering the original signals requires sufficient observations.
However, when the original signals are assumed to be sparse, the $L_1$-norm minimization problem can lead to the exact answer even from a limited number of observations~\cite{CandesTao2006,Donoho2006_2}.
By use of the $L_1$ norm, we can extract several significant influencers from a vast number of combinations hidden in dataset.
Therefore the $L_1$ norm is often utilized in various realms in the data analysis~\cite{Lasso1994,Yamanaka2015,Ohzeki2015,Ezaki2017,Otsuki2017,Shinaoka2017}.
In terms of the Hamiltonian estimation, we do not necessarily need a number of pairs of the spin configurations and energy values to infer the number of sparse interactions.

In the present study, we formulate the Hamiltonian estimation as the $L_1$-norm minimization problem and propose a method to estimate an unknown cost function utilizing $L_1$-norm minimization.
In addition, we demonstrate its theoretical assessment.
As our formulation is of the same form of the well-known problem in CS, it can be evaluated using a replica method~\cite{Nishimori2001,Dotsenko2000}, which is a sophisticated tool in the field of statistical mechanics~\cite{Kabashima2009,Ganguli2010,Krzakala2012,Rangan2009,Xu2013,Ohzeki2015deep} that theoretically guarantees the performance of reconstruction for various problems.
To verify our analysis, several numerical experiments are performed, in which we estimate unknown coupling constants in the Ising Hamiltonian by employing the alternating direction method of multipliers~\cite{Boyd2011} (ADMM).
Then, the theoretical limits of the reconstruction are compared with the numerical results.
We show that our analysis via the replica method is validated through the numerical experiments.

The remainder of the present paper is organized as follows.
\sref{sec:formulation} shows the formulation of Hamiltonian estimation problem and presents the method for solving it by using the CS framework.
\sref{sec:replica} shows the analysis of the typical performance of our proposed method through the replica method.
In \sref{sec:experiments}, we show the numerical results of the estimation of coupling constants using the ADMM, and compare it with our analytical result discussed in \sref{sec:replica}.
In \sref{sec:conclusion}, we summarize our results and discuss several future outlooks.

\section{Hamiltonian estimation}
\label{sec:formulation}
We first explain the problem setting in our study.
From only a set of pairs of the energy values and spin configurations, we estimate the coupling constants of the Hamiltonian.
In several unfortunate cases, we cannot formulate directly the cost function of the optimization problem.
We assume that it is hard to obtain the value of the cost function (energy values).
For example, for finding the optimal choice of drugs for curing diseases, we need to confirm the effect of the combination.
In general, the results cannot be obtained immediately.
Therefore, we have only a small number of pairs of the cost function and the choice of the drugs in terms of binary variables (chosen or not).
In such a case, if we can find the cost function from only a limited number of ``hints'', we can find the optimal solution from direct optimization of the estimated Ising Hamiltonian.

We formulate the above problem as follows.
Let us consider a general full-connect Ising system with $N$ sites.
The Ising spins $\sigma_i\in\{-1,+1\}$ are located at sites $i\in V$, where $V$ is a set of sites in the system.
The coupling constants between spins are defined as $J_{ij}$ for any two combinations of sites $i$ and $j$ in $V$.
We assume that the interaction between spins are symmetric and no self-interaction occurs.
The Hamiltonian is assumed to be expressed as
\begin{align}
  H(\bm{\sigma}) := - \frac{1}{N}\sum_{i < j} J_{ij} \sigma_i \sigma_j.
  \label{eq:Hamiltonian}
\end{align}
In addition, the Hamiltonian is called an energy function in the field of information science.
Its observed value is called energy, especially in physics.
We then obtain a set of $M$ ($<N(N-1)/2$) values of the energy $\bm{E} := (E^{(1)}, E^{(2)}, \cdots, E^{(M)})^{\mathrm{T}}$ and an observation matrix of spin configurations $S$, which is constructed by the given spin configurations.
Here, $S$ is an $M \times N(N-1)/2$ matrix expressed as
\begin{widetext}
\begin{align*}
  S :=
    \begin{pmatrix}
      \sigma_1^{(1)}\sigma_2^{(1)} & \cdots & \sigma_1^{(1)}\sigma_N^{(1)} & \sigma_2^{(1)}\sigma_3^{(1)} & \cdots & \sigma_{N-1}^{(1)}\sigma_{N}^{(1)}\\
      \sigma_1^{(2)}\sigma_2^{(2)} & \cdots & \sigma_1^{(2)}\sigma_N^{(2)} & \sigma_2^{(2)}\sigma_3^{(2)} & \cdots & \sigma_{N-1}^{(2)}\sigma_{N}^{(2)}\\
      \vdots & \cdots & \vdots & \vdots & \cdots & \vdots\\
      \sigma_1^{(M)}\sigma_2^{(M)} & \cdots & \sigma_1^{(M)}\sigma_N^{(M)} & \sigma_2^{(M)}\sigma_3^{(M)} & \cdots & \sigma_{N-1}^{(M)}\sigma_{N}^{(M)}
    \end{pmatrix}
  .
\end{align*}
\end{widetext}
We assume that a set of $N(N-1)/2$ original coupling constants is expressed as
$\bm{J}^0 := (J_{12}^0, \cdots, J_{1N}^0, J_{23}^0, \cdots, J_{N-1,N}^0)^{\mathrm{T}}$,
and a set of energies $\bm{E}$ is expressed as
\begin{align}
  \bm{E} := - \frac{1}{N} S \bm{J}^0.
  \label{eq:energy}
\end{align}
In general, the case that the number of unknown variables $\bm{J}^0$ is larger than the number of linear equations as in \eref{eq:energy}, is an underdetermined system.
When we simply solve \eref{eq:energy}, we cannot obtain a unique solution.
Hence, a method that can adequately estimate the unknown $\bm{J}$ only from the given $M$ pairs of energies and spin configurations is desired.

To attain the above-mentioned objective, we utilize the concept of CS.
We impose an assumption of sparsity on $\bm{J}^0$.
In particular, we assume that the probability distribution of $J_{ij}^0$ is
\begin{align}
  P(J_{ij}^0) := (1 - \rho)\delta(J_{ij}^0) + \rho \mathcal{N}(0,1),
  \label{eq:posterior}
\end{align}
where $\delta(x)$ denotes the Dirac delta function,
$\rho = 2K / N(N-1)$ ($0\leq\rho\leq 1$) denotes the density of the nonzero components in the vector $\bm{J}^0$,
and $\mathcal{N}(0,1)$ denotes the Gaussian distribution with a vanishing mean and unit variance.
Note that we do not know which component in $\bm{J}^0$ is nonzero or zero.

We then formulate the estimate of $\bm{J}^0$ as the $L_1$-norm minimization with respect to $\bm{J}:= (J_{12}, \cdots, J_{1N}, J_{23}, \cdots, J_{N-1,N})^{\mathrm{T}}$ based on the CS concept.
The $L_1$-norm minimization in our problem setting is formulated as follows:
\begin{align}
  \min_{\bm{J}} \|\bm{J}\|_1 \ \ \mathrm{subject\ to}\ \ \bm{E} = -\frac{1}{N} S \bm{J}.
  \label{eq:minl1norm}
\end{align}
where $\|\bm{J}\|_1:=\sum_{i<j}|J_{ij}|$ denotes the $L_1$ norm with respect to $\bm{J}$.
The solution of \eref{eq:minl1norm} under the constraint yields an adequate estimate of the unknown $\bm{J}^{0}$ under several conditions with respect to $\rho$ and $\alpha = 2M/N(N-1)$.
The problem in \eref{eq:minl1norm} can be solved using various optimization algorithms,
such as the coordinate descent~\cite{Wright2015},
fast iterative shrinkage-thresholding algorithm~\cite{FISTA2009},
augmented Lagrangian method~\cite{Hestenes1969,Powell1976}, or ADMM~\cite{Gabay1976,Boyd2011}.

\section{Replica analysis}
\label{sec:replica}
In \sref{sec:formulation}, we formulated the $L_1$-norm minimization problem in our problem setting.
In this study, we emphasize on how the estimation performance obtained through our method depends on $\rho$ and $\alpha$.
In the present section, we investigate the typical behavior of our method proposed in \sref{sec:formulation}.
To assess the proposed method, we employ the replica method~\cite{Nishimori2001,Dotsenko2000}.
The replica method has been developed in the field of statistical physics and is frequently used to analyze the free energy, which is defined through a partition function as the normalization constant of the distribution and is difficult to calculate analytically in most cases.
The replica method gives an analytical representation of the configurational average of a partition function instead of the partition function itself based on the self-averaging property.

The typical behavior of the estimation of the coupling constants described in \eref{eq:minl1norm} can be investigated by analyzing the partition function of the following distribution by using the replica method:
\begin{align}
  P(\bm{J}\mid\bm{E}) = \frac{1}{Z_{\beta}(\bm{E})}P(\bm{E}\mid\bm{J})P(\bm{J}),
\end{align}
where $Z_{\beta}(\bm{E})$ is the partition function of this distribution.
The logarithm of $Z_{\beta}(\bm{E})$ is an important quantity, namely, the free energy, and is defined as
\begin{align}
  f = & - \lim_{\beta \rightarrow \infty}\lim_{N\rightarrow \infty}\frac{1}{\beta N^2}[\ln Z_{\beta}(\bm{E})]_{S,\bm{J}^0}\nonumber\\
  = & - \lim_{\beta \rightarrow \infty}\lim_{n\rightarrow 0}\frac{\partial}{\partial n}\lim_{N \rightarrow \infty}\frac{1}{\beta N^2}\ln [Z_{\beta}^n(\bm{E})]_{S,\bm{J}^0},
  \label{eq:freeenergy}
\end{align}
where $[\cdots]_{S,\bm{J}^0}$ denotes the configurational average with respect to $S$ and $\bm{J}^0$.
According to the procedure of the general replica analysis~\cite{Kabashima2009}, we evaluate $[Z_{\beta}^n (\bm{E})]_{S, \bm{J}^0}$ in \eref{eq:freeenergy} for $n\in\mathbb{N}$.
Furthermore, $[Z_{\beta}^n (\bm{E})]_{S, \bm{J}^0}$ is performed with an analytical continuation for $n\in\mathbb{R}$ after completing the computation of the partition function into a closed form and obtaining a function form of $n$.

By using the expression $E^{(\mu)} := - \sum_{i<j}J_{ij}^0 \sigma_i^{(\mu)} \sigma_j^{(\mu)}/N$,
$[ Z_{\beta}^n(\bm{E})]_{S,\bm{J}^0}$ is expressed as
\begin{align}
  &[Z_{\beta}^n(\bm{E})]_{S,\bm{J}^0} \nonumber\\
  &\quad = \left[\int\prod_{a=1}^{n}d\bm{J}^a \prod_{\mu =1}^M
  \delta(S(\bm{J}^a - \bm{J}^0))\exp\left(-\beta \| \bm{J}^a \|_1 \right)\right]_{S,\bm{J}^0},
  \label{eq:Z_beta}
\end{align}
where $\bm{J}^a$ denotes the $a$-th replicated vector of the coupling constants, and $\|\bm{J}^a\|_1$ denotes the $L_1$ norm of $\bm{J}^a$.
Here, we assumed $P(\bm{J}^a)$ $\forall a \in \{1,2,\cdots,n\}$ to be the Laplace distribution: $P(\bm{J}^a)\propto\exp(-\beta\|\bm{J}^a\|_1)$.
Note that \eref{eq:Z_beta} is only valid $\forall n \in \mathbb{N}$.
To obtain an analytical expression for a part of \eref{eq:Z_beta}, $\left[\prod_{a=1}^n \delta (S(\bm{J}^a - \bm{J}^0))\right]_{S}$, we define the following quantity:
\begin{align}
  u_{\mu}^a := \frac{1}{N}\sum_{i<j}(J_{ij}^a - J_{ij}^0)\sigma_i^{(\mu)}\sigma_j^{(\mu)},
\end{align}
where $a = 0, 1, 2, \cdots, n$ is the index of the replica.
Here, we assume that
\begin{align}
  \big\langle \sigma_i^{(\mu)} \sigma_j^{(\mu)} \big\rangle_{\bm{\sigma}^{(\mu)}} &= \big\langle \sigma_i^{(\mu)} \big\rangle_{\bm{\sigma}^{(\mu)}} \big\langle \sigma_j^{(\mu)} \big\rangle_{\bm{\sigma}^{(\mu)}} = c^2,
  \label{eq:sigmacondition1}\\
  \big\langle \sigma_i^{(\mu)} \sigma_j^{(\mu)} \sigma_k^{(\mu)} \sigma_l^{(\mu)} \big\rangle_{\bm{\sigma}^{(\mu)}} &= \Big\langle \big(\sigma_i^{(\mu)}\big)^4 \Big\rangle_{\bm{\sigma}^{(\mu)}} \nonumber\\
  & = \left\{
  \begin{array}{ll}
  c^4 & (i\neq j\neq k\neq l)\\
  1 & (i=k, j=l; i=l, j=k),
  \label{eq:sigmacondition2}
  \end{array}
  \right.
\end{align}
where $\langle \cdots \rangle_{\bm{x}}$ denotes $\sum_{\bm{x}}(\cdots) P(\bm{x})$,
and $c\ (-1 \leq c \leq 1)$ is a constant.
Then, we consider the case in which $c = 0$, i.e., $\langle\sigma_i^{(\mu)}\rangle_{\bm{\sigma}^{(\mu)}} = 0$.
\eref{eq:sigmacondition1} and \eref{eq:sigmacondition2} are rewritten as
\begin{align}
  \big\langle \sigma_i^{(\mu)} \sigma_j^{(\mu)} \big\rangle_{\bm{\sigma}^{(\mu)}} &= 0,
  \label{eq:sigmacondition3}\\
  \big\langle \sigma_i^{(\mu)} \sigma_j^{(\mu)} \sigma_i^{(\mu)} \sigma_j^{(\mu)} \big\rangle_{\bm{\sigma}^{(\mu)}} &= \delta_{ik}\delta_{jl} + \delta_{il}\delta_{jk},
  \label{eq:sigmacondition4}
\end{align}
respectively.
The assumptions in Eqs.~\eqref{eq:sigmacondition1}--\eqref{eq:sigmacondition4} are unique characteristics in the models with Ising variables.
From the assumptions in Eqs.~\eqref{eq:sigmacondition1}--\eqref{eq:sigmacondition4}, $\langle u_{\mu}^a \rangle_{\bm{\sigma}^{(\mu)}} = 0$ is obtained.
Similarly, we obtain $\langle u_{\mu}^a u_{\mu}^b \rangle_{\bm{\sigma}^{(\mu)}} = 2 (q^{00}-q^{0a}-q^{a0}+q^{ab})$, where
\begin{align}
  q^{ab} := \frac{1}{N^2} \sum_{i<j} J_{ij}^a J_{ij}^b
  \label{eq:qab}
\end{align}
are the order parameters with respect to the coupling constants.
Herein, we assume the following replica symmetric ansatz to extend \eref{eq:Z_beta} to $n\in\mathbb{R}$:
\begin{align}
  q^{ab} = \left\{
  \begin{array}{ll}
  \rho & (a=b=0)\\
  m & (a\neq 0, b=0; a=0, b\neq 0)\\
  q & (a\neq b, a\neq 0, b\neq 0)\\
  Q & (a=b, a\neq 0, b\neq 0).
  \end{array}
  \right.
  \label{eq:RSansatz}
\end{align}
By considering \eref{eq:RSansatz}, $u_{\mu}^a$ can be represented as ($n+1$) multivariate Gaussian random variables with a vanishing mean and covariance $\langle u_{\mu}^a u_{\mu}^b \rangle_{\bm{\sigma}^{(\mu)}} = 2(\rho - 2m + Q)$ for $a = b$ and $\langle u_{\mu}^a u_{\mu}^b \rangle_{\bm{\sigma}^{(\mu)}} = 2(\rho - 2m + q)$ for $a \neq b$.
$u_{\mu}^a$ in \eref{eq:umu} can be expressed in the new form
\begin{align}
  u_{\mu}^a = \sqrt{Q-q}s_a + \sqrt{\rho-2m+q}t,
  \label{eq:umu}
\end{align}
where $s_a$ and $t$ are the Gaussian random variables with vanishing mean and unit variance.
By using the representation in \eref{eq:umu}, the following expression is obtained:
\begin{align}
  \left[\;\prod_{a=1}^n \delta(S(\bm{J}^a - \bm{J}^0))\right]_{S}&= \lim_{\tau\rightarrow 0} \prod_{a=1}^{n} \prod_{\mu=1}^{M}\frac{1}{\sqrt{2\pi\tau}}\left[ \exp\left(-\frac{(u_{\mu}^a)^2}{2\tau}\right)\right]_{\bm{u}}\nonumber\\
  &\approx \exp\frac{nM}{2}\left(\ln(Q-q) - \frac{\rho-2m+q}{Q-q}\right),
  \label{eq:mean_u}
\end{align}
where $[\cdots]_{\bm{u}}$ denotes $\int Dt \int Ds_a (\cdots)$, and $\int Dx := \int dx \exp(-x^2/2)/ \sqrt{2\pi}$.
Here, we used $\exp(nx) = 1 + nx + O(x^2)$ and $\delta(x) = \lim_{\tau\rightarrow 0} \exp(-x^2/2\tau) / \sqrt{2\pi\tau}$.
See \aref{sec:appendixA} for the detailed form of $[Z_{\beta}(\bm{E})]_{S,\bm{J}^0}$ introduced in the results up to \eref{eq:mean_u} and several other expressions.

From the replica symmetric ansatz in \eref{eq:RSansatz} and the saddle-point approximations in \aref{sec:appendixB}, the right-hand side of \eref{eq:freeenergy} can be explicitly expressed as
\begin{align}
  f &= \extr_{Q,\chi,m,\tilde{Q},\tilde{\chi},\tilde{m}} \Biggl\{ \frac{\alpha(\rho-2m+Q)}{2\chi} -\frac{1}{2}(Q\tilde{Q} - \chi\tilde{\chi}) + m\tilde{m} \nonumber\\
  & \quad + (1-\rho)\int D t\; \Phi(t;\tilde{Q}, \tilde{\chi},0) + \rho\int D t\; \Phi(t; \tilde{Q},\tilde{\chi},\tilde{m}) \Biggr\},
  \label{eq:extr_freeenergy}
\end{align}
where $\tilde{Q}, \tilde{\chi}$, and $\tilde{m}$ are auxiliary parameters for introducing the integral expressions of the Kronecker delta for the replica symmetric solutions $Q, \chi$, and $m$, respectively.
Here, we define
\begin{align}
  &\Phi(t; \tilde{Q},\tilde{\chi},\tilde{m}) \nonumber\\
  &\quad := - \frac{1}{2\tilde{Q}}\left(\left|\sqrt{\tilde{\chi}+\tilde{m}^2}t\right|-1\right)^2\;\Theta\left(\left|\sqrt{\tilde{\chi}+\tilde{m}^2}t\right|-1\right),
  \label{eq:Phi}
\end{align}
where
\begin{align*}
  \Theta(x) := \left\{
  \begin{array}{ll}
  1 & (x > 0)\\
  0 & (x \leq 0).
  \end{array}
  \right.
\end{align*}
See \aref{sec:appendixB} for an overview of a more detailed derivation of \eref{eq:extr_freeenergy}.

The extremization of \eref{eq:extr_freeenergy} yields the following saddle-point equations:
\begin{align}
  \begin{split}
    \tilde{Q} &= \frac{\alpha}{\chi},\ \ \tilde{\chi} = \frac{\alpha}{\chi^2}(\rho-2m+Q),\ \ \tilde{m} = \frac{\alpha}{\chi},\\
    Q &= \frac{2(1-\rho)}{\tilde{Q}^2}G(\tilde{\chi}) + \frac{2\rho}{\tilde{Q}^2} G(\tilde{\chi}+\tilde{m}),\\
    \chi &= \frac{2(1-\rho)}{\tilde{Q}}H\left(\frac{1}{\sqrt{\tilde{\chi}}}\right) + \frac{2\rho}{\tilde{Q}} H\left(\frac{1}{\sqrt{\tilde{\chi}+\tilde{m}}}\right),\\
    m &= \frac{2\rho\tilde{m}}{\tilde{Q}}H\left(\frac{1}{\sqrt{\tilde{\chi}+\tilde{m}}}\right),
    \label{eq:saddlepoint}
  \end{split}
\end{align}
where we define $H(x) := \int_x^{\infty} Dt$ and $G(x) := (x+1)H\left(1/\sqrt{x}\right) - \sqrt{x/2\pi}\exp\left(-1/2x\right)$.
The equations in \eref{eq:extr_freeenergy} and Eqs.~\eqref{eq:saddlepoint} coincide with the results by Kabashima et al.~\cite{Kabashima2009} and Ganguli and Sompolinsky~\cite{Ganguli2010}.
Therefore, the stability condition for the successful estimation by using the $L_1$-norm minimization in our problem setting is expressed as
\begin{align}
  \alpha > 2(1-\rho)H\left(\frac{1}{\sqrt{\tilde{\chi}}}\right) + \rho,
  \label{eq:reconstructionlimit}
\end{align}
which is also similar to the results by Kabashima et al.~\cite{Kabashima2009} and Ganguli and Sompolinsky~\cite{Ganguli2010}.

We can also obtain an analytical expression for the typical value of the mean squared error (MSE), which is often used as a performance index in various fields.
By using the extremum solutions in Eqs.~\eqref{eq:saddlepoint}, the MSE is expressed as
\begin{align}
  [\mathrm{MSE}]_{S,\bm{J}^0} &= \frac{2}{N(N-1)} \left[ \big\langle \|\bm{J}-\bm{J}^0\|_2^2 \big\rangle_{\bm{J}\mid\bm{E}}^{\beta\rightarrow \infty} \right]_{S,\bm{J}^0} \nonumber\\
  &= \rho-2m+Q,
  \label{eq:MSE}
\end{align}
where $\|\bm{x}\|_2$ denotes the $L_2$ norm $\|\bm{x}\|_2 := (\sum_{i}x_i^2)^{1/2}$ for $\bm{x} = \{x_i\}$,
and $\langle \cdots \rangle_{\bm{x}}^{\beta\rightarrow\infty}$ denotes the average with respect to $P(\bm{x})$ as $\beta\rightarrow\infty$.

\begin{figure*}[htb]
\centering
\includegraphics[width=9.5cm]{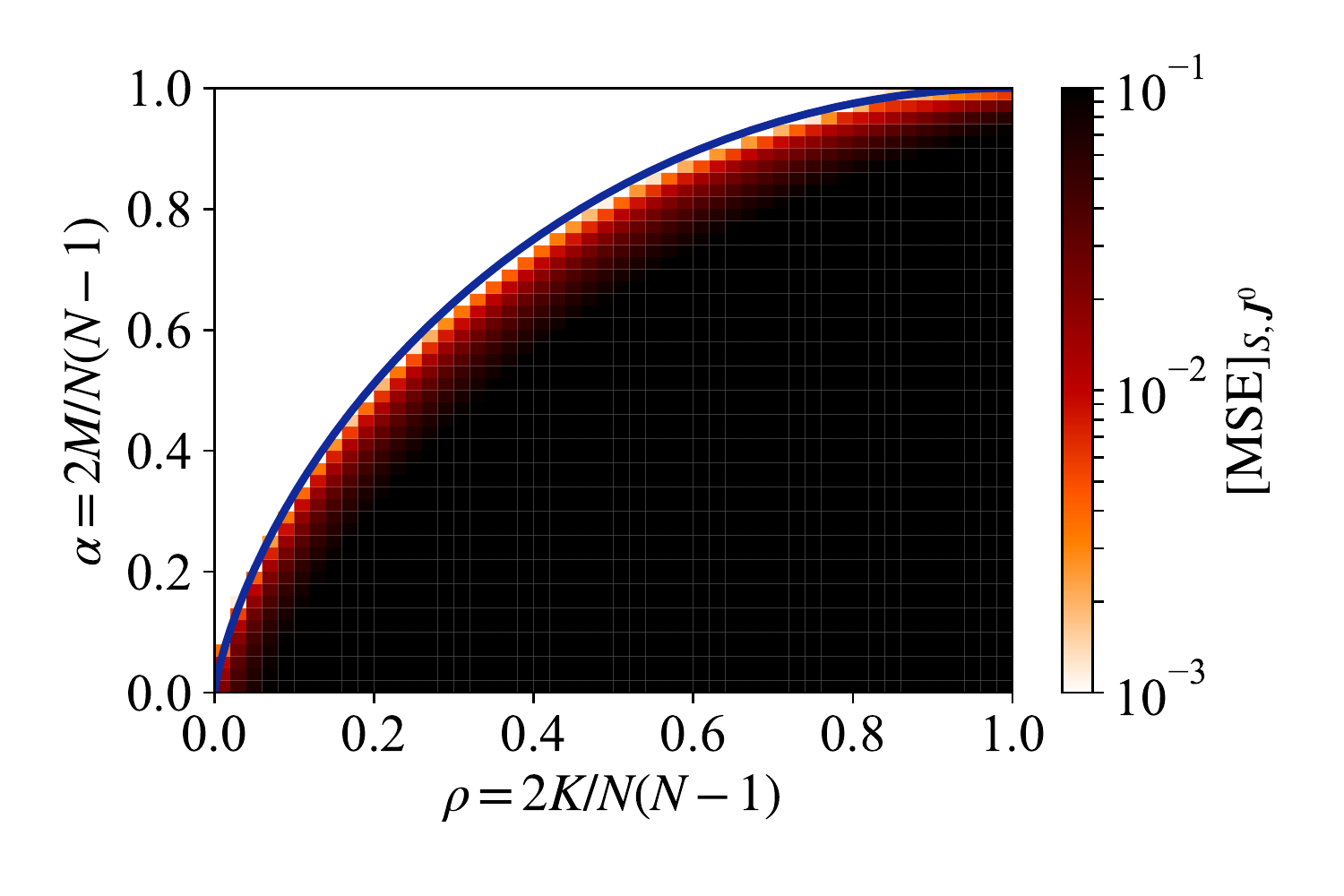}
\caption{
(Color online) Typical reconstruction limit of the $L_1$-norm minimization obtained from the replica analysis.
The blue curve shows the boundary $\textstyle\alpha = 2(1-\rho)H(1/\sqrt{\tilde{\chi}}) + \rho$.
The region colored white to black via red shows the values of the MSE in \eref{eq:MSE}, which indicates whether the estimation succeeds or fails.
The white and black blocks represent regions of successful and failed estimations, respectively.
}
\label{fig:replica}
\end{figure*}

Figure \ref{fig:replica} shows the result of our analysis in this section.
The blue curve shows the theoretical reconstruction limit of the true coupling constants estimated through the $L_1$-norm minimization, expressed as $\alpha = 2(1-\rho)H(1/\sqrt{\tilde{\chi}})+\rho$.
The heat map in \fref{fig:replica} shows the expectations of the MSE described in \eref{eq:MSE}.
The region colored white to black via red shows the values of the MSE in \eref{eq:MSE}, indicating whether the estimation succeeds or fails.
The expectation of the MSE is calculated using $Q$ and $m$, which are obtained by updating $Q,m,\chi,\tilde{Q},\tilde{m}$, and $\tilde{\chi}$ using Eqs.~\eqref{eq:saddlepoint}, until $(x^{k+1}-x^k)$ reaches a value less than $10^{-10}$ for $\forall x\in\{Q,m,\chi,\tilde{Q},\tilde{m},\tilde{\chi}\}$.
The values of $\rho$ and $\alpha$ are varied in steps of $0.02$ in the ranges of $0.02\leq\rho\leq 1$ and $0.02\leq\alpha\leq 1$.
The averages of the MSEs of $100$ trials for each $\rho$ and $\alpha$ are plotted in each block of the heat map, containing $50\times 50 = 2500$ blocks in the map.
The white blocks represent regions where the estimation succeeds.
Here, we determined that the estimation \textit{succeeds} when the MSE in \eref{eq:MSE} reaches a value less than $10^{-3}$.
In contrast, the black blocks indicate regions where estimation fails.
The typical performance can be achieved by adequately solving the $L_1$-norm minimization problem.

\section{Numerical verifications}
\label{sec:experiments}

In this section, we numerically demonstrate the performance of our proposed method.
In the experiments, we use an Ising model whose the Hamiltonian is described in \eref{eq:Hamiltonian}.
We employ the ADMM~\cite{Boyd2011} to solve the $L_1$-norm minimization in \eref{eq:minl1norm} with the constraint described in \eref{eq:posterior}.
This solves the optimization problem $\min_{\bm{x}} f(\bm{x})$ by splitting variables $\bm{x}\in\mathbb{R}^{d}$ into $\bm{x}\in\mathbb{R}^{d}$ and $\bm{z}\in\mathbb{R}^{d}$, where $d$ denotes the number of dimensions of an original set of variables, and $\min_{\bm{x},\bm{z}} \{ f(\bm{z}) + g(\bm{x})\}$ is solved subject to
$\bm{x}-\bm{z}=\bm{0}$, where $f(\bm{z})$ and $g(\bm{x})$ are assumed to be convex functions.
The optimization is attained by minimizing the augmented Lagrangian, $L(\bm{x},\bm{z};\bm{\lambda}) = f(\bm{z}) + g(\bm{x}) + \bm{\lambda}^{\mathrm{T}}(\bm{x}-\bm{z}) + r \|\bm{x}-\bm{z}\|_2^2 / 2$,
where $\bm{\lambda}:=\{\lambda_i\mid i=1,2,\cdots,d\}$ denotes a set of Lagrange multipliers, and $r$ denotes a penalty constant.
We can obtain optimal $\bm{x},\bm{z}$ and $\bm{\lambda}$ by iterating the following equations:
$\bm{x}^{k+1} = \argmin_{\bm{x}}L(\bm{x},\bm{z}^k;\bm{\lambda}^k)$, $\bm{z}^{k+1} = \argmin_{\bm{z}}L(\bm{x}^{k+1},\bm{z};\bm{\lambda}^k)$,
and $\bm{\lambda}^{k+1} = \bm{\lambda}^{k} + r(\bm{x}^{k+1}-\bm{z}^{k+1})$, alternately,
where $k$ is the number of the iteration steps.

Based on the procedure by Boyd et al.~\cite{Boyd2011},
the following optimization problem is considered in our experiments:
\begin{align}
  \min_{\bm{J},\bm{K}}\left\{ \|\bm{K}\|_1 + \bm{\lambda}^{\mathrm{T}}\left(\bm{E}+\frac{1}{N}S\bm{J}\right) \right\}\ \ \mathrm{subject\ to}\ \ \bm{J}-\bm{K} = \bm{0},
  \label{eq:optproblem}
\end{align}
where $\bm{K}$ is a vector consisting of $N(N-1)/2$ components,
and $\bm{\lambda}$ is a vector consisting of $N(N-1)/2$ Lagrange multipliers for the constraint described in \eref{eq:minl1norm}.
To solve the problem in \eref{eq:optproblem},
we minimize the augmented Lagrangian for \eref{eq:optproblem} described as
\begin{align}
  L(\bm{J},\bm{K};\bm{\Lambda}) = \|\bm{K}\|_1 + r\bm{\Lambda}^{\mathrm{T}}\left(\bm{E}+\frac{1}{N}S\bm{J}\right) + \frac{r}{2}\left\|\bm{J}-\bm{K}+\bm{\Lambda}\right\|_2^2,
\end{align}
where $\bm{\Lambda} = \bm{\lambda}/r$.
Optimal $\bm{J},\bm{K}$ and $\bm{\Lambda}$ are obtained by iteratively calculating the following equations:
\begin{align}
  \bm{J}^{k+1} &= \left(I - S^{\mathrm{T}}(SS^{\mathrm{T}})^{-1}S\right)(\bm{K}^k-\bm{\Lambda}^k)-NS^{\mathrm{T}}(SS^{\mathrm{T}})^{-1}\bm{E},
  \label{eq:update1}\\
  \bm{K}^{k+1} &= \mathrm{ST}_{1/r}(\bm{J}^{k+1}+\bm{\Lambda}^{k}),
  \label{eq:update2}\\
  \bm{\Lambda}^{k+1} &= \bm{\Lambda}^k + \bm{J}^{k+1} - \bm{K}^{k+1},
  \label{eq:update3}
\end{align}
where $I$ denotes a $N(N-1)/2 \times N(N-1)/2$ identity matrix,
and $\mathrm{ST}_{y}(x)$ denotes the soft-thresholding function~\cite{STDonoho,STDaubechies} defined as
\begin{align*}
  \mathrm{ST}_{y}(x) = \left\{
  \begin{array}{ll}
    x - y & (x > y)\\
    0 & (-y \leq x \leq y)\\
    x + y & (x < -y).
  \end{array}
  \right.
\end{align*}

\begin{figure*}[htb]
\centering
\includegraphics[width=14.7cm]{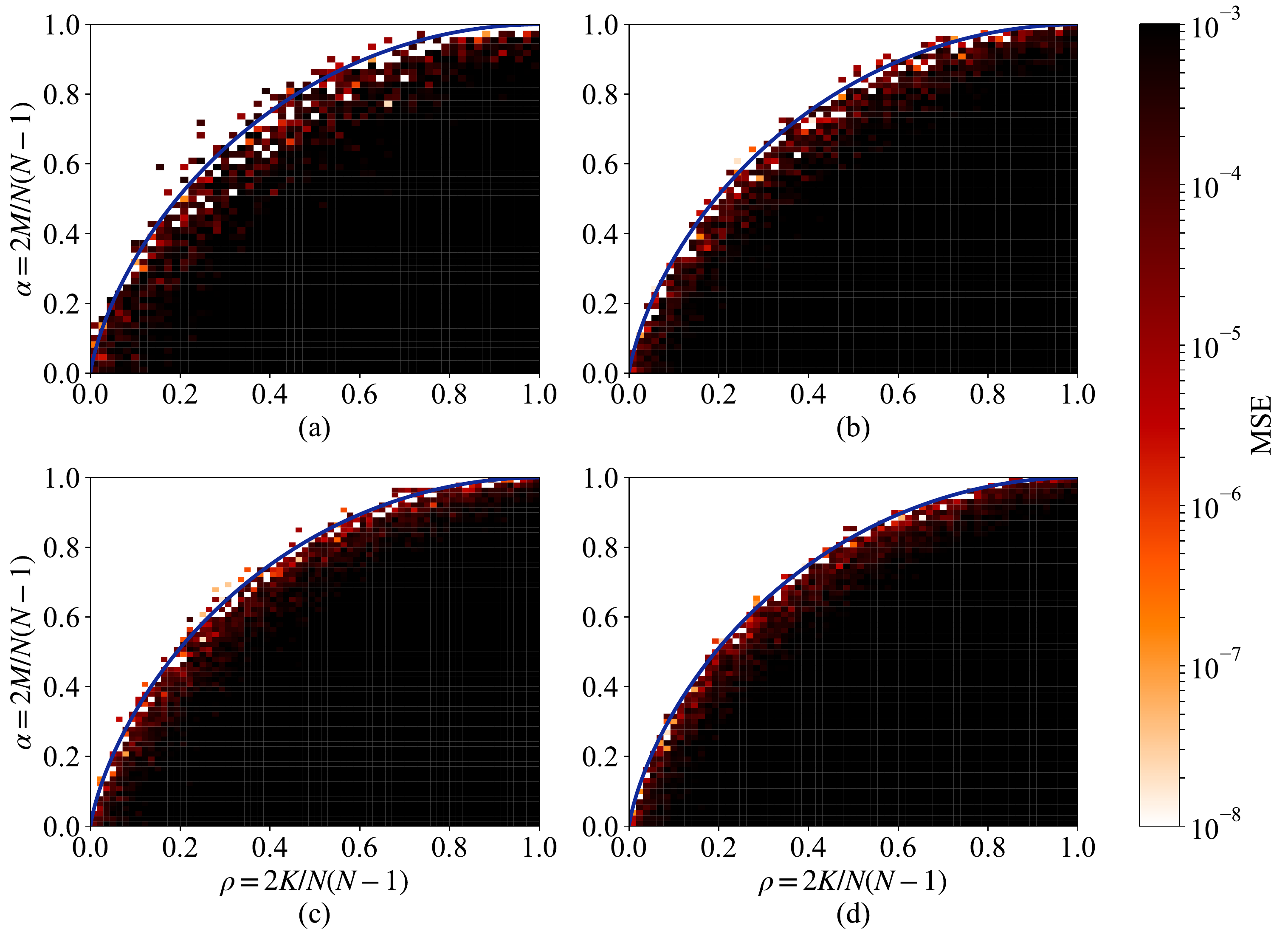}
\caption{
(Color online) Comparisons of the analytical reconstruction limit $\alpha = 2(1-\rho)H(1/\sqrt{\tilde{\chi}}) + \rho$ with the MSEs obtained from the numerical simulations through the ADMM.
The blue curve lines show the reconstruction limit in \eref{eq:reconstructionlimit} in each plot.
The white blocks represent the success of the estimation, while the black blocks represent the failure.
(a) MSEs versus $\rho$ and $\alpha$ when $N=11$, $N(N-1)/2=55$.
(b) MSEs versus $\rho$ and $\alpha$ when $N=16$, $N(N-1)/2=120$.
(c) MSEs versus $\rho$ and $\alpha$ when $N=21$, $N(N-1)/2=210$.
(d) MSEs versus $\rho$ and $\alpha$ when $N=26$, $N(N-1)/2=325$.
}
\label{fig:ADMM}
\end{figure*}

Figure \ref{fig:ADMM} shows the comparison of the analytical reconstruction limit in
\eref{eq:reconstructionlimit} with the MSEs obtained through the ADMM.
The true coupling constants $\bm{J}^0$ were generated from the distribution in \eref{eq:posterior}.
$\sigma_i^{(\mu)}$ for $\forall i \in V$ and $\forall \mu \in M$ are independently generated from the uniform distribution $P(\sigma_i^{(\mu)} = -1) = P(\sigma_i^{(\mu)} = +1) = 1/2$.
To measure the performances, we used the MSE between true coupling constants and estimated coupling constants defined as $\mathrm{MSE} = 2\sum_{i<j}(J_{ij}-J_{ij}^0) /N(N-1)$.
The blue curve lines at each plot in \fref{fig:ADMM} show the theoretical reconstruction limits of true coupling constants obtained in \sref{sec:replica}.
The color of each block in the heat map represents the averaged values of the MSEs over 100 trials calculated through the ADMM.
Eqs.~\eqref{eq:update1}--\eqref{eq:update3} were iteratively updated up to 500 times per trial.
For all the experiments in \fref{fig:ADMM}, $r=1$.
We determined that the estimation succeeded when the value of the averaged MSE reaches less than $10^{-8}$.
Otherwise, the estimation was said to have failed.
Figure \ref{fig:ADMM}~(a) shows the averaged MSEs when $N=11$, $N(N-1)/2=55$;
these are described in $55\times 55=3025$ blocks, \fref{fig:ADMM}~(b) demonstates the averaged MSEs when $N=16$, $N(N-1)/2=120$;
these are described in $60\times 60=3600$ blocks.
Figure \ref{fig:ADMM}~(c) shows the averaged MSEs when $N=21$, $N(N-1)/2=210$;
these are described in $70\times 70=4900$ blocks and \fref{fig:ADMM}~(d) describes the averaged MSEs when $N=26$, $N(N-1)/2=325$;
these are described in $65\times 65=4225$ blocks.
Through the experiments, we showed that the black region converges to the inside $\alpha \leq 2(1-\rho)H(1/\sqrt{\tilde{\chi}})+\rho$ with the system size increases.
This behavior is natural as the replica method premises that the system size is infinitely large.
Therefore, we obtained several empirical results, and thus our analysis can be considered to be valid.

\section{Conclusions and discussions}
\label{sec:conclusion}

In this paper, we proposed a method for estimating unknown coupling constants in an Ising model by using the CS technique.
Moreover, we analytically investigated the performance of our proposed method using the replica analysis.
Finally, we compared our analytical result of the replica analysis and the numerical results obtained through the ADMM, and showed that our analytical results mostly coincide with the numerical results.

The proposed method can be regarded as a general method for sparse representation of the cost functions of various optimization problems.
Owing to the limitations of hardware designs, the implementation of an optimization problem is not easy task.
When we embed the Ising model in actual machines, the constraints with respect to their structures must be considered.
For instance, the D-Wave machine is restricted to a chimera graph, which has a sparse number of edges.
Thus, various techniques are desirable for making the structure of the optimization problem sparse without losing the important features of the original one.
The construction of such a methodology can contribute to the extension of the practical uses of the machines in several fields.

The point of the analysis in \sref{sec:replica} was the assumption for the components of the observation matrix described in Eqs.~\eqref{eq:sigmacondition1}--\eqref{eq:sigmacondition4}.
We are also interested in the performance of the estimation when the observation matrix has another constraint.
For instance, we can consider the case where we can ask for the value of the cost function for the oracle by sending the spin configuration.
Then, the spin configuration is assumed to be completely random as in the present study.
However, it is more efficient to send a spin configuration that is different from the previously sent configurations.
In other words, orthogonal observation should be valuable for inferring the Hamiltonian from the pairs of spin configurations and the value of the cost function.
In our future studies, we will assess the theoretical performance when $S$ changes and the design the structure of $S$ to attain a better estimate of the Hamiltonian using the replica method or other statistical-mechanical analyses, such as cavity method~\cite{cavity2003,cavity2009}.
In the present study, we focus on the estimation of the Hamiltonian itself.
Depending on the problem setting, the resulting ground state is important.
While estimating the Hamiltonian, we may find the ground-state spin configuration.
Away from the procedure based on CS, we will report the direct manipulation of optimization while estimating the Hamiltonian only from a small number of observations on the cost function in the future.

\begin{acknowledgments}
The authors would like to thank Shu Tanaka, Shun Kataoka and Yuya Seki for insightful comments and suggestions.
One of the authors (C.T.) was partially supported by Grant-in-Aid for JSPS Fellows from the Japan Society for the Promotion of Science Grant (No. JP17J03081) and JST-CREST for Japan Science and Technology Agency (No.JPMJCR1402).
One of the authors (M.O.) was partially supported by Inamori Foundation, KAKENHI No.15H03699, No.16H04382, and No.16K13849, ImPACT Program of Council for Science, Technology and Innovation (Cabinet Office, Government of Japan) and JST-START.
\end{acknowledgments}

\onecolumngrid
\appendix
\section{Detailed form of $[Z_{\beta}^{n} (\bm{E})]_{S,\bm{J}^0}$ at equation (\ref{eq:mean_u})}
\label{sec:appendixA}
In this appendix, we supplement the derivation of $[Z_{\beta}^n (\bm{E})]_{S,\bm{J}^0}$ before applying the saddle-point approximations by using several formulas.
To specify the introduction of the order of parameters $q^{ab}$ in \eref{eq:qab},
we insert $1 = \prod_{a,b}\int dq^{ab} \delta(q^{ab}-\sum_{i<j}J_{ij}^a J_{ij}^b/N^2)$ into \eref{eq:Z_beta}.
$[Z_{\beta}^n (\bm{E})]_{S,\bm{J}^0}$ can be rewritten as
\begin{align}
  [Z_{\beta}^n (\bm{E})]_{S,\bm{J}^0} = \prod_{a,b}\int dq^{ab}\prod_{a=1}^{n}\prod_{i<j}\int dJ_{ij}^a \left[ \left[\;\prod_{a=1}^n \delta(S(\bm{J}^a - \bm{J}^0))\right]_{S} \prod_{a,b}\delta\left( q^{ab}-\frac{1}{N^2}\sum_{i<j}J_{ij}^a J_{ij}^b \right) \exp\left( -\beta\sum_{i<j}\left|J_{ij}^a\right| \right) \right]_{\bm{J}^0},
  \label{eq:Z_re1}
\end{align}
where an analytical expression for a part of the right-hand side was given in \eref{eq:mean_u}.
We then consider rewriting $\prod_{a,b}\int dq^{ab} \delta(q^{ab}-\sum_{i<j}J_{ij}^a J_{ij}^b/N^2)$ in more detail.
From the replica symmetric ansatz in \eref{eq:RSansatz}, we obtain
\begin{align*}
&\prod_{a,b}\int dq^{ab} \delta\left(q^{ab}-\frac{1}{N^2}\sum_{i<j}J_{ij}^a J_{ij}^b\right)\nonumber\\
&\quad= \prod_{a,b}\int dQ\int dq \int dm\;\delta\left(Q-\frac{1}{N^2}\sum_{i<j}J_{ij}^a J_{ij}^a\right)\delta\left(q-\frac{1}{N^2}\sum_{i<j}J_{ij}^a J_{ij}^b\right)\delta\left(m-\frac{1}{N^2}\sum_{i<j}J_{ij}^a J_{ij}^0\right).
\end{align*}
Furthermore, the following expressions are introduced:
\begin{align}
  \delta  \biggl(Q-\frac{1}{N^2} \sum_{i<j} J_{ij}^a J_{ij}^a\biggr) &= \int d\tilde{Q}\exp\left( \frac{\tilde{Q}}{2}\biggl(N^2 Q - \sum_{i<j} J_{ij}^a J_{ij}^a\biggr) \right),
  \label{eq:delta1}\\
  \delta  \biggl(q-\frac{1}{N^2} \sum_{i<j} J_{ij}^a J_{ij}^b\biggr) &= \int d\tilde{q}\exp\left( -\frac{\tilde{q}}{2}\biggl(N^2 q - \sum_{i<j} J_{ij}^a J_{ij}^b\biggr) \right),
  \label{eq:delta2}\\
  \delta  \biggl(m-\frac{1}{N^2} \sum_{i<j} J_{ij}^a J_{ij}^0\biggr) &= \int d\tilde{m}\exp\Biggl( -\tilde{m}\biggl(N^2 m - \sum_{i<j} J_{ij}^a J_{ij}^0\biggr) \Biggr),
  \label{eq:delta3}
 \end{align}
where $\tilde{Q},\tilde{q}$, and $\tilde{m}$ are auxiliary parameters for introducing the above integral representations of the Kronecker delta functions.
By employing the expressions in \eref{eq:mean_u} and Eqs.~\eqref{eq:delta1}--\eqref{eq:delta3} and the Hubbard--Stratonovich transformation
\begin{align*}
  \prod_{a\neq b} \exp\left(\frac{\tilde{q}}{2}J_{ij}^a J_{ij}^b\right) = \int Dt \prod_{a=1}^n \exp\left(\sqrt{\tilde{q}} J_{ij}^a t - \frac{\tilde{q}}{2}\left(J_{ij}^a\right)^2\right),
\end{align*}
\eref{eq:Z_re1} can be expressed as
\begin{align}
  [Z_{\beta}^n (\bm{E})]_{S,\bm{J}^0} &= \int dQ \int dq \int dm \int d\tilde{Q} \int d\tilde{q} \int d\tilde{m}\nonumber\\
  &\quad \times \exp\left( n N^2 \left( \frac{1}{2}\left(\tilde{Q}Q + \tilde{q}q \right) - \tilde{m}m + \frac{\alpha}{2}\left( \ln(Q-q) - \frac{\rho-2m+q}{Q-q} \right) \right) \right)\nonumber\\
  &\quad \times \left[ \int Dt \prod_{a=1}^{n} \prod_{i<j} \int d J_{ij}^{a} \exp \left( -\frac{\tilde{Q}+\tilde{q}}{2}\left(J_{ij}^{a}\right)^2 + \left( \tilde{m}J_{ij}^{0} + \sqrt{\tilde{q}}t \right) J_{ij}^{a} - \beta \left|J_{ij}^{a}\right| \right) \right]_{\bm{J}^0}.
  \label{eq:Z_re2}
\end{align}
Note that we approximated $M$ by using $\alpha N^2$ in \eref{eq:mean_u}.

\section{Detailed derivation of equation \textbf{(\ref{eq:extr_freeenergy})}}
\label{sec:appendixB}
We developed an expression for $[Z_{\beta}^n (\bm{E})]_{S,\bm{J}^0}$ in \aref{sec:appendixA}.
However, \eref{eq:Z_re2} still has many integrations.
In this appendix, we explain the procedure of the saddle-point method~\cite{Nishimori2011}.

First, we consider the saddle-point approximations for the integrations with respect to $Q,q,m,\tilde{Q},\tilde{q}$, and $\tilde{m}$.
A part of \eref{eq:Z_re2} can be rewritten as
\begin{align}
  &\left[ \int Dt \prod_{a=1}^{n} \prod_{i<j} \int d J_{ij}^{a} \exp \left( -\frac{\tilde{Q}+\tilde{q}}{2}\left(J_{ij}^{a}\right)^2 + \left( \tilde{m}J_{ij}^{0} + \sqrt{\tilde{q}}t \right) J_{ij}^{a} - \beta \left|J_{ij}^{a}\right| \right) \right]_{\bm{J}^0}\nonumber\\
  &\quad \approx nN^2 \left[ \int Dt \int dJ \exp \left( -\frac{\tilde{Q}+\tilde{q}}{2}J^2 + \left(\tilde{m} J^0 + \sqrt{\tilde{q}}t\right)J - \beta |J| \right) \right]_{J^0},
  \label{eq:ptZ}
\end{align}
where $J$ and $J^0$ are univariate parameters.
We substitute \eref{eq:ptZ} into \eref{eq:Z_re2} and apply the saddle-point approximation $\int dx \exp f(x) \approx \exp f(x^*)$ to $Q,q,m,\tilde{Q},\tilde{q}$, and $\tilde{m}$.
We obtain
\begin{align}
  [Z_{\beta}^n (\bm{E})]_{S,\bm{J}^0} \approx & \exp\left( n N^2 \Biggl( \frac{1}{2}\left(\tilde{Q}Q + \tilde{q}q \right) - \tilde{m}m + \frac{\alpha}{2}\left( \ln(Q-q) - \frac{\rho-2m+q}{Q-q} \right)\right.\nonumber\\
  &\qquad\qquad\quad \times \left.\left. \left[ \int Dt \int dJ \exp \left( -\frac{\tilde{Q}+\tilde{q}}{2}J^2 + \left(\tilde{m} J^0 + \sqrt{\tilde{q}}t\right)J - \beta |J| \right) \right]_{J^0}\right)\right),
  \label{eq:Z_re3}
\end{align}
where the parameters at the saddle points are denoted as $Q^*\rightarrow Q,\ q^*\rightarrow q,\ m^*\rightarrow m,\ \tilde{Q}^*\rightarrow \tilde{Q},\ \tilde{q}*\rightarrow \tilde{q}$, and $\tilde{m}^* \rightarrow \tilde{m}$.

Here, let us consider the effect of the temperature on the parameters.
We assume that $\tilde{Q}+\tilde{q}\rightarrow\beta\tilde{Q}$, $Q-q\rightarrow\chi/\beta$, $\tilde{m}\rightarrow\beta\tilde{m}$, and $\tilde{q}\rightarrow\beta^2\tilde{\chi}$ as $\beta\rightarrow \infty$.
From these, we obtain
\begin{align}
  &\left[ \int Dt \int dJ \exp \left( -\frac{\tilde{Q}+\tilde{q}}{2}J^2 + \left(\tilde{m} J^0 + \sqrt{\tilde{q}}t\right)J - \beta |J| \right) \right]_{J^0} \rightarrow \beta \left( (1-\rho)\int Dt \Phi(t;\tilde{Q},\tilde{\chi},0) + \rho \int Dt \Phi(t;\tilde{Q},\tilde{\chi},\tilde{m}) \right),
\end{align}
where the transform $\tilde{m}J^0 + \sqrt{\tilde{\chi}}t \rightarrow \sqrt{\tilde{\chi} + \tilde{m}^2}t$ and the saddle-point approximation $\int dJ \exp f(J) \approx \exp f(J^*)$ are applied.

By using $\exp(nx)=1+nx+O(x^2)$ and $\ln(1+x)=x+O(x^2)$ and substituting these expressions into \eref{eq:freeenergy}, the analytical expression of free energy in \eref{eq:extr_freeenergy} is finally obtained.

\twocolumngrid
\bibliographystyle{jpsj}
\bibliography{cs}

\end{document}